\documentclass[aps,prd,showpacs,showkeys,amsmath,amssymb,
twocolumn,
floatfix,
superscriptaddress,
nofootinbib
]{revtex4}
\pdfoutput=1

\usepackage{graphicx}
\usepackage{multirow}
\usepackage{longtable}
\usepackage{lscape}
\usepackage{rotating}
\usepackage{color}

\usepackage{amsmath,amstext,amsfonts,amsbsy,amssymb,amscd,bbm,epsfig,lscape}
\usepackage{hyperref}

\newcommand{\ie}{\textit{i.e.}}
\newcommand{\eg}{\textit{e.g.}}

\newcommand{\be}{\begin{equation}}
\newcommand{\ee}{\end{equation}}
\newcommand{\bea}{\begin{eqnarray}}
\newcommand{\eea}{\end{eqnarray}}

\newcommand{\epsee}{\epsilon_{ee}}
\newcommand{\epsem}{\epsilon_{e\mu}}
\newcommand{\epset}{\epsilon_{e\tau}}

\allowdisplaybreaks[4] % for page break

\begin{document}

\begin{titlepage}

\renewcommand{\thefootnote}{\alph{footnote}}

\vspace*{-1.7cm}

\begin{flushright}{\small FERMILAB-PUB-16-115-T}\end{flushright}

\vspace*{0.2cm}

\title{Generalized mass ordering degeneracy in neutrino oscillation experiments}

\author{Pilar Coloma}
\affiliation{Fermi National Accelerator Laboratory, P.O. Box 500, Batavia, IL 60510, USA }
\author{Thomas Schwetz}
\affiliation{Institut f\"ur Kernphysik, Karlsruhe Institute
  of Technology (KIT), 76021 Karlsruhe, Germany}

\begin{abstract}
  We consider the impact of neutral-current (NC) non-standard neutrino
  interactions (NSI) on the determination of the neutrino mass
  ordering. We show that in presence of NSI there is an exact
  degeneracy which makes it impossible to determine the neutrino mass
  ordering and the octant of the solar mixing angle $\theta_{12}$ at
  oscillation experiments. The degeneracy holds at the probability
  level and for arbitrary matter density profiles, and hence, solar,
  atmospheric, reactor, and accelerator neutrino experiments are
  affected simultaneously. The degeneracy requires order-one corrections from NSI to the NC
neutrino--quark interaction and can be tested in 
neutrino--nucleus NC scattering experiments.
\end{abstract}

\maketitle

\end{titlepage}

%%%%%%%%%%%%%%%%%%%%%%%%%%
\section{Introduction}
%%%%%%%%%%%%%%%%%%%%%%%%%%
Neutrino oscillation physics has entered the precision era. Present
data determines all three leptonic mixing angles 
and the absolute value of the two mass-squared differences with few
percent precision~\cite{Gonzalez-Garcia:2014bfa}. Crucial goals of
future oscillation experiments are ($a$) the determination of 
the neutrino mass ordering and the CP-violating 
phase $\delta$, and ($b$) establishing the robustness of  
three-flavour oscillations with respect to physics beyond
the Standard Model (SM). In the present work we show that those two items
are intimately related. We consider the hypothesis that additional interactions
affect the neutrino sector, beyond the SM weak interaction~\cite{Wolfenstein:1977ue,Valle:1987gv,Guzzo:1991hi}, see~\cite{Ohlsson:2012kf,Miranda:2015dra} for recent reviews. We will show that, for
a certain choice of these non-standard interactions (NSI), the
determination of the neutrino mass ordering---one of the main goals of
upcoming oscillation experiments~\cite{Messier:2013sfa,
  Acciarri:2015uup, Aartsen:2014oha, Djurcic:2015vqa,
  Ahmed:2015jtv}---becomes impossible, due to an exact degeneracy in
the evolution equation governing neutrino oscillations in matter. 

The paper is structured as follows. In Sec.~\ref{sec:NSI} we introduce the NSI framework and the notation used in the rest of the paper. Section~\ref{sec:deg} shows the origin of the degeneracy and how it can be realized in both vacuum and matter regimes. In Sec.~\ref{sec:osc} we explain how the degeneracy affects neutrino oscillation data, while in Sec.~\ref{sec:scattering} we explore the possible combination with neutrino scattering data to try to remove the degeneracy. Finally, our conclusions are summarized in Sec.~\ref{sec:conclusions}. 

%%%%%%%%%%%%%%%%%%%%%%%%%%
\section{Non-Standard Interactions in neutrino propagation}
\label{sec:NSI}
%%%%%%%%%%%%%%%%%%%%%%%%%%
Three-flavour neutrino evolution in an arbitrary matter potential is
described by the Schroedinger equation
\begin{equation} \label{eq:schroedinger}
  i \frac{d}{dx} \Psi = H(x) \Psi \,,
\end{equation}
where $\Psi$ is a vector of the flavour amplitudes, $\Psi = (a_e,
a_\mu, a_\tau)^T$, and $H(x) = H_{\rm vac} + H_{\rm mat}(x)$. The
Hamiltonian describing evolution in vacuum is 
\begin{equation}
  H_{\rm vac} = U \text{diag}(0, \Delta_{21}, \Delta_{31}) U^\dagger \,,
\end{equation}
with $\Delta_{ij} = \Delta m^2_{ij} / (2E_\nu)$, where $\Delta
m^2_{ij} \equiv m^2_i - m^2_j$ stands for the neutrino mass-squared
difference, and $E_\nu$ is the neutrino energy. From neutrino oscillation data, we know that 
$|\Delta m^2_{31}| \approx |\Delta m^2_{32}| \approx 30 \Delta
m^2_{21}$. The neutrino mass ordering is parametrized by
the sign of the larger mass-squared difference, with normal ordering
(NO) corresponding to $\Delta m^2_{31} > 0$ and inverted ordering (IO) to $\Delta m^2_{31} < 0$. 
The sign of $\Delta m^2_{21}$ by convention
is chosen positive. The standard parametrization for the leptonic
mixing matrix is $U = O_{23} U_{13} O_{12}$, where $O_{ij}$ ($U_{ij}$)
denotes a real (complex) rotation in the $ij$ sector, with mixing angle
$\theta_{ij}$. Here we find it convenient to use an equivalent
parametrization, where we put the complex phase $\delta$ in the 12
rotation, such that $U = O_{23} O_{13} U_{12}$. After subtracting a term proportional to the unit matrix, the vacuum Hamiltonian becomes
\begin{equation}\label{eq:Hvac1}
  H_{\rm vac} = O_{23}O_{13} \left(\begin{array}{cc}
   H^{(2)} & 0 \\
  0 & \Delta_{31} - \frac{\Delta_{21}}{2}
\end{array} \right) O_{13}^T O_{23}^T \, ,
\end{equation}
with the 12 block given by 
\begin{equation}\label{eq:Hvac2}
  H^{(2)} = \frac{\Delta_{21}}{2 }\left(\begin{array}{cc}
    -\cos 2\theta_{12}  & \sin 2 \theta_{12} e^{i\delta}  \\
    \sin 2 \theta_{12} e^{-i\delta}  & \cos 2\theta_{12} \\
\end{array} \right) \,.
\end{equation}

Let us consider now the presence of neutral-current (NC) NSI in the
form of dimension-6 four-fermion operators, which may contribute to
the effective potential in matter in $H_{\rm mat}$. We follow the
notation of \cite{Gonzalez-Garcia:2013usa}, for a recent review see
\eg~\cite{Miranda:2015dra}. NSI are described by the Lagrangian
\begin{equation}\label{eq:L}
  \mathcal{L}_{\rm NSI} = -2\sqrt{2} G_F \epsilon_{\alpha\beta}^f
  (\overline\nu_{\alpha L} \gamma^\mu \nu_{\beta L}) (\overline f \gamma_\mu f) \,,
\end{equation}
where, $\alpha, \beta = e,\mu,\tau$, and $f$ denotes a fermion present
in the background medium. The parameter $\epsilon_{\alpha\beta}^f$
parametrizes the strength of the new interaction with respect to the
Fermi constant $G_F$. Hermiticity requires that
$\epsilon_{\alpha\beta}^f = (\epsilon_{\beta\alpha}^f)^*$. Note that
we restrict to vector interactions, since we are interested in the
contribution to the effective matter potential.  In generic models of new physics NSI parameters are expected to be small. However, examples of viable gauge models leading to $\epsilon_{\alpha\beta}^{u,d} \sim \mathcal{O}(1)$ can be found in \cite{Farzan:2015doa, Farzan:2015hkd} (see also \cite{Miranda:2015dra} for a discussion of NSI models). 

The matter part of the Hamiltonian is then obtained as
\begin{align}
%  H_{\rm mat} &= \sqrt{2} G_F N_e (x) [ {\rm diag}(1,0,0) + (\epsilon_{\alpha\beta})] \, , \label{eq:Hmat}\\
  H_{\rm mat} &= \sqrt{2} G_F N_e (x) 
  \left( \begin{array}{ccc} 1 + \epsilon_{ee}& \epsilon_{e\mu} & \epsilon_{e\tau} \\
  \epsilon_{e\mu}^* & \epsilon_{\mu\mu} & \epsilon_{\mu\tau} \\
  \epsilon_{e\tau}^* & \epsilon_{\mu\tau}^* & \epsilon_{\tau\tau}  \end{array} \right)   , \label{eq:Hmat}\\[3mm]
  \epsilon_{\alpha\beta} &= \sum_{f=e,u,d} Y_f(x) \epsilon_{\alpha\beta}^f \, , \label{eq:eps}
\end{align}
with $Y_f(x) \equiv N_f(x)/N_e(x)$, $N_f(x)$ being the density of
fermion $f$ along the neutrino path. This implies that the effective NSI
parameters $\epsilon_{\alpha\beta}$ may depend on $x$.  The ``1'' in the $ee$ entry in eq.~\eqref{eq:Hmat} corresponds to the
standard matter potential \cite{Wolfenstein:1977ue, Mikheev:1986gs}.
For neutral matter, the densities of electrons and protons are
equal. Thus, the relative densities of up and down quarks are
\begin{equation}\label{eq:Yq}
    Y_u (x) = 2 + Y_n (x) \,,\quad
    Y_d (x) = 1 + 2Y_n (x) \,,  
\end{equation}
where $Y_n(x)$ is the relative neutron density along the neutrino
path.  Below we will use the notation $\epsilon_{\alpha\beta}^\oplus$ and $\epsilon_{\alpha\beta}^\odot$ to indicate when the
$\epsilon_{\alpha\beta}$ refer to the specific matter composition of
the Earth or the Sun, respectively.

%%%%%%%%%%%%%%%%%%%%%%%%%%
\section{The generalized mass ordering degeneracy}
\label{sec:deg}
%%%%%%%%%%%%%%%%%%%%%%%%%%
Let us consider first the vacuum part of the Hamiltonian, $H_{\rm vac}$
defined in eqs.~\eqref{eq:Hvac1} and \eqref{eq:Hvac2}. It is easy to show that
the transformation
\begin{equation} \label{eq:vac}
  \begin{array}{l}
    \Delta m^2_{31} \to -\Delta m^2_{31} + \Delta m^2_{21} = -\Delta m^2_{32} \,,\\
    \sin\theta_{12} \leftrightarrow \cos\theta_{12} \,,\\
    \delta \to \pi - \delta 
  \end{array}
\end{equation}
implies that $H_{\rm vac} \to - H_{\rm vac}^*$.  Inserting this into
eq.~\eqref{eq:schroedinger} and taking the complex conjugate we
recover exactly the same evolution equation, when we take into account
that complex conjugation of the amplitudes ($\Psi \to \Psi^*$) is
irrelevant, as only moduli of flavour amplitudes are
observable.\footnote{The invariance of the evolution under the
  transformation $H \to -H^*$ is a consequence of CPT invariance. It
  has been noted in the context of NSI in \cite{GonzalezGarcia:2011my}
  and applied in some limiting cases, see also
  \cite{Gonzalez-Garcia:2013usa}.} This proves that the transformation
\eqref{eq:vac} leaves the three-flavour evolution in vacuum invariant.

Note that this transformation corresponds to a complete inversion of
the neutrino mass spectrum. The transformation $\Delta m^2_{31} \leftrightarrow -\Delta m^ 2_{32}$
exchanges NO and IO, while changing the octant of $\theta_{12}$
exchanges the amount of $\nu_e$ present in 
$\nu_1$ and $\nu_2$. We denote the effect of the transformation
\eqref{eq:vac} as ``flipping'' the mass spectrum. The corresponding
degeneracy is known in limiting cases, for instance, the so-called
mass ordering degeneracy in the context of long-baseline experiments
\cite{Minakata:2001qm}. It is manifest also in the exact
expression for the three-flavour $\nu_e$ survival-probability $P_{ee}$
in vacuum, relevant for medium-baseline reactor experiments~\cite{Bakhti:2014pva}.

It is clear that for a non-zero standard matter effect,
eq.~\eqref{eq:Hmat} with $\epsilon_{\alpha\beta} = 0$, the
transformation \eqref{eq:vac} no longer leaves the evolution
invariant, since $H_{\rm mat}$ remains constant. The matter effect in the 13-sector
is the basis of the mass ordering determination in long-baseline
accelerator \cite{Messier:2013sfa, Acciarri:2015uup} and 
atmospheric neutrino \cite{Djurcic:2015vqa, Ahmed:2015jtv}
experiments. Moreover, the observation of the MSW~\cite{Wolfenstein:1977ue, Mikheev:1986gs}
matter resonance in the Sun requires that $\theta_{12} < 45^\circ$,
which forbids the transformation in the second line of
eq.~\eqref{eq:vac}. This allows, in principle, for the determination of the
mass ordering via a precise measurement of $P_{ee}$ in vacuum~\cite{Petcov:2001sy}, as
intended for instance by the JUNO collaboration
\cite{Aartsen:2014oha}. 

However, if in addition to the transformation \eqref{eq:vac},
it is also possible to transform $H_{\rm mat} \to -H_{\rm mat}^*$, then the full Hamiltonian
including matter would transform as $H \to -H^*$, leaving the
evolution equation invariant. This can be achieved in presence of
NSI, supplementing the transformation \eqref{eq:vac} with~\cite{Bakhti:2014pva}
\begin{equation}\label{eq:trf-mat}
  \begin{array}{l}
    \epsilon_{ee} \to -\epsilon_{ee} - 2 \,,\\
    \epsilon_{\alpha\beta} \to - \epsilon_{\alpha\beta}^* \quad (\alpha\beta \neq ee) \,.
  \end{array}
\end{equation}
The transformation of $\epsilon_{ee}$ is crucial to change the sign of
the $ee$ element of $H_{\rm mat}$ including the standard matter
effect. Note that eq.~\eqref{eq:trf-mat} depends on the
parametrization used for $H_{\rm vac}$ in eq.~\eqref{eq:Hvac1}. If
the standard parametrization with $U = O_{23}
U_{13} O_{12}$ was used instead, then we would obtain $\epsem \to \epsem^*$, $\epset
\to \epset^*$, and all other $\epsilon_{\alpha\beta}$ transforming as
in eq.~\eqref{eq:trf-mat}.

Since in general the $\epsilon_{\alpha\beta}$ are dependent on the
neutron density, the degeneracy can be broken in principle by
comparing experiments in matter with different neutron densities, or
in configurations where the neutron density changes significantly
along the neutrino path. However, one can choose couplings such that
NSI with neutrons are zero and take place only with protons and/or
electrons, by choosing $\epsilon_{\alpha\beta}^q$ proportional to the
quark electric charge, \ie, $\epsilon_{\alpha\beta}^u =
-2\epsilon_{\alpha\beta}^d$. In this situation the
$\epsilon_{\alpha\beta}$ are always independent of $x$, the
degeneracy is complete and cannot be broken by any combination of
neutrino oscillation experiments.

Let us illustrate the degeneracy by the following example: assume that
there are no NSI in Nature. Then we can fit data from any neutrino
oscillation experiment either with standard oscillations and the
correct spectrum, or equally well with a flipped spectrum and
$\epsilon_{ee} = -2$. For
\begin{equation}\label{eq:exact}
\epsilon_{ee}^u = -4/3 \,,\quad 
\epsilon_{ee}^d = 2/3 
\end{equation}
we obtain $\epsilon_{ee} = -2$ independent of the neutron density, and
hence the degeneracy will be perfect, irrespective of the matter
environment. 

%%%%%%%%%%%%%%%%%%%%%%%%%%
\section{Impact of the degeneracy at oscillation experiments}
\label{sec:osc} 
%%%%%%%%%%%%%%%%%%%%%%%%%%

A manifestations of this result is
the so-called LMA-dark solution for solar neutrinos
\cite{Miranda:2004nb}, which corresponds to a fit to solar neutrino
data with $\theta_{12} > 45^\circ$ (``dark octant'') and values of
$\epsilon_{ee}^{u,d} \simeq -1$. In the Sun the neutron fraction $Y_n$
drops from about 1/2 in the centre to about 1/6 at the border of the
solar core. From eqs.~\eqref{eq:Yq} and \eqref{eq:eps} follows,
that for $\epsilon_{ee}^{u,d} \simeq -1$ we obtain $\epsee^\odot \simeq
-2$, close to the value needed for the generalized mass ordering
degeneracy. In \cite{Gonzalez-Garcia:2013usa} a recent analysis of
solar neutrino data has been performed, assuming either NSI with up or
down quarks. In this case $Y_n$ does not drop out of
$\epsilon_{\alpha\beta}$ defined in eq.~\eqref{eq:eps}, and hence the
condition $\epsee = -2$ cannot be fulfilled along the whole
neutrino path in the Sun. Therefore, the degeneracy is not perfect. In
\cite{Gonzalez-Garcia:2013usa} the $\Delta\chi^2$ of the LMA-dark
solution is nearly zero for NSI on up quarks and $\lesssim 2$ for down
quarks. While the sign of $\Delta m_{31}^2$ is irrelevant for solar
neutrino phenomenology, it has been realised in \cite{Bakhti:2014pva},
that the $\sin\theta_{12} \leftrightarrow \cos\theta_{12}$ ambiguity
introduced by the LMA-dark solution leads to a mass ordering ambiguity
in the planned reactor experiment JUNO \cite{Aartsen:2014oha}. This is
a manifestation of the generalized degeneracy discussed above.

\begin{figure}[t!]
  \includegraphics[scale=0.54]{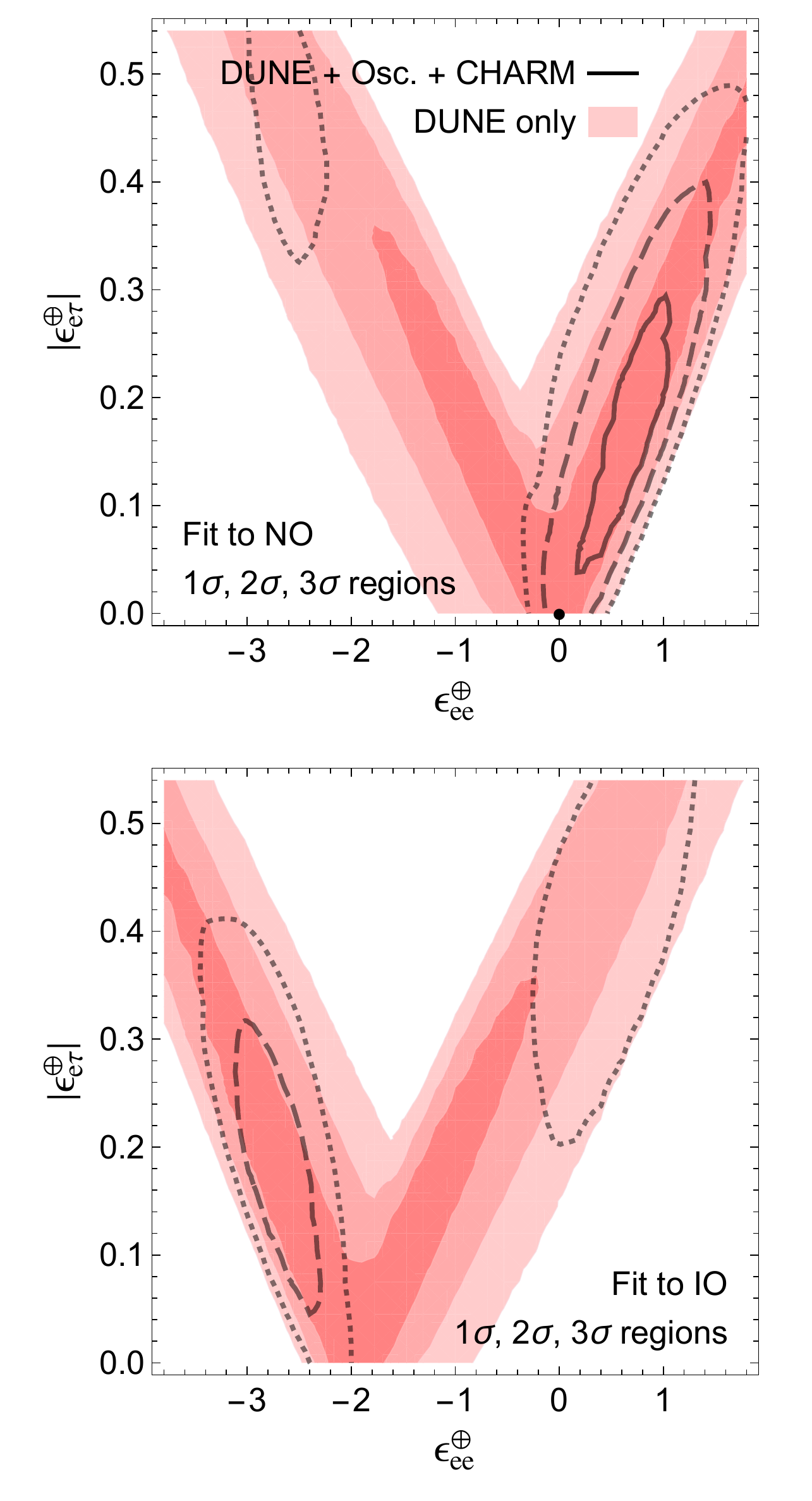}
  \caption{Results from a fit to simulated data for DUNE. We assume a
    true NO and no NSI, and perform a fit allowing for non-zero
    values of $\epsee$ and $\epset$. In the upper panel we fit with
    the correct mass spectrum, while in the lower panel we adopt IO
    and exchange $\sin\theta_{12} \leftrightarrow \cos\theta_{12}$. The shaded
    regions correspond to DUNE alone, whereas the contour curves
    include the constraints from global oscillation data
    \cite{Gonzalez-Garcia:2013usa} and from the CHARM experiment
    \cite{Dorenbosch:1986tb} on the NC cross section. We marginalize
    over $\Delta m^2_{31}$, $\delta$, $\theta_{23}$ and the complex phase of $\epset$.}
\label{fig:Vplot}
\end{figure}

As another example, we will now
demonstrate the impact of the generalized degeneracy on the
sensitivity of the long-baseline Deep Underground Neutrino Experiment (DUNE)
\cite{Acciarri:2015uup} to the mass ordering.
In absence of NSI, the DUNE experiment would be able to reject the
wrong hypothesis for the mass ordering with a significance above $\sim
5\sigma$ regardless of the true value of
$\delta$~\cite{Acciarri:2015uup}.
We calculate expected data for NO, $\delta
= 40^\circ$, $\sin^ 2\theta_{12} = 0.3$, and no NSI. The simulation is performed
using GLoBES \cite{Huber:2007ji}; the simulation details are the same as in Ref.~\cite{Coloma:2015kiu}. Then these artificial data are
fitted by allowing for the simultaneous presence of $\epsilon_{ee}$ and
$\epsilon_{e\tau}$, while all other NSI parameters are set to zero,
for simplicity. Results are shown by the shaded regions in
Fig.~\ref{fig:Vplot}. The lower panel confirms the perfect degeneracy
of the flipped mass spectrum at $\epsee^\oplus = -2$ and $\epset = 0$, with 
$\Delta \chi^2 = 0$ with respect to the true best fit point.
In both panels of Fig.~\ref{fig:Vplot} we observe also a strong correlation
between $\epsee$ and $\epset$, see \cite{Coloma:2015kiu}. Therefore, while
the degeneracy is exact for $(\epsee,\epset) = (-2,0)$,
it is recovered to a good accuracy for nonzero values of $\epset$
as long as $|\epset| \simeq 0.2 |\epsee+2 |$. The importance of $\epsee$
and $\epset$ for the mass ordering determination in DUNE has been pointed
out recently in \cite{Liao:2016hsa}.

%%%%%%%%%%%%%%%%%%
%%%%%%%%%%%%%%%%%%

%%%%%%%%%%%%%%%%%%%%%%%%%%
\section{Combination with neutrino scattering data} 
\label{sec:scattering}
%%%%%%%%%%%%%%%%%%%%%%%%%%

Since the generalized degeneracy is
exact and holds for any oscillation experiment, the only way to break
it are non-oscillation experiments. Indeed, operators of the type in
eq.~\eqref{eq:L} contribute to the neutral current (NC) neutrino
scattering cross section. Unfortunately, data on electron neutrino NC
scattering is scarce. A relevant constraint on the parameters of
interest to us comes from the historical CHARM
experiment~\cite{Dorenbosch:1986tb}, which has measured the quantity
$R_e = 0.406 \pm 0.140$, where $R_e$ is ratio of the electron neutrino
plus antineutrino NC cross sections to the corresponding charged
current ones. In presence of NSI we have
$R_e = {\tilde g}_L^2 + {\tilde g}_R^2$, where \cite{Escrihuela:2009up}
\begin{equation}
  {\tilde g}_P^2 = \sum_{q=u,d} \left[
    \left(g_P^q + \frac{\epsee^q}{2} \right)^2 +
    \frac{|\epsem^q|^2 + |\epset^q|^2}{4}
    \right] \,,
\end{equation}
with $P=L,R$, and $g_P^q$ are the SM NC couplings. We have
included only the vector-like NSI parameters. Note that the CHARM constraint is somewhat model dependent, 
since it would not apply if the mediator particle responsible for the NSI is much lighter than the momentum transfer
in CHARM (typically of several tens of GeV) \cite{Farzan:2015doa}.

%%%%%%%%%%%%%%%%%%
\begin{figure}[t!]
  \includegraphics[scale=0.5]{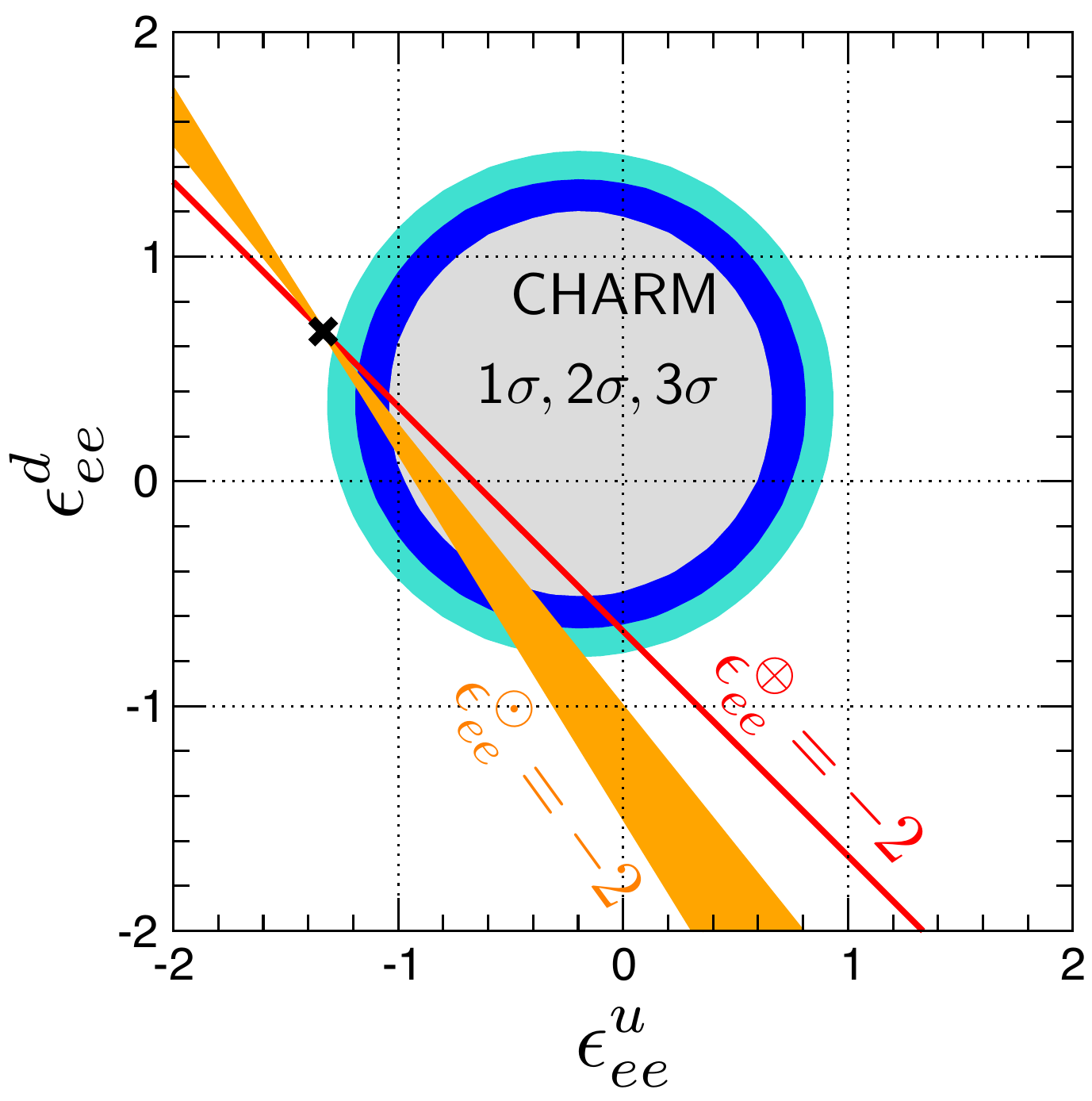}
  \caption{Allowed region at $1,2,3\sigma$
    (2~d.o.f.) from CHARM~\cite{Dorenbosch:1986tb} (assuming NSI from a heavy mediator) in the plane of $\epsee^u$ and $\epsee^d$, for $\epsem^{u,d} = \epset^{u,d}  = 0$. The cross
    corresponds to the point of exact degeneracy,
    eq.~\eqref{eq:exact}. The diagonal lines indicate the parameters
    for which $\epsee = -2$ in Earth and solar matter.}
\label{fig:charm}
\end{figure}
%%%%%%%%%%%%%%%%%%

Assuming that the CHARM bound applies, we follow Ref.~\cite{Escrihuela:2009up} and we show in Fig.~\ref{fig:charm} the
allowed region for $\epsee^u$ and $\epsee^d$ from the CHARM data.  The
point from eq.~\eqref{eq:exact}, corresponding to perfect degeneracy
for any matter profile, is indicated by the cross in the figure.
We observe that it is excluded by CHARM data: for
this point we predict $R_e \approx 0.956$, which disagrees with the CHARM
experimental value at $3.9\sigma$. The diagonal
lines in the figure indicate the parameters for which $\epsee = -2$ in Earth and
solar matter. We use that in Earth matter, $Y_n \approx 1.05$, and for
the Sun we show the spread induced by $Y_n = 1/2 \to 1/6$. For
neutrino trajectories in the Earth, the generalized degeneracy holds
along the line indicated in the plot. However, since the degeneracy
for the Sun appears for slightly different values of $\epsee^{u,d}$
there is the potential to break it by the combination. Indeed, for the
LMA-dark solution for NSI either on up or down quarks we have
$\epsee^{u,d} \approx -1$~\cite{Gonzalez-Garcia:2013usa}. From
Fig.~\ref{fig:charm} we see that $(\epsee^{u},\epsee^{d}) \approx (0,-1)$ is in strong
disagreement with CHARM, while $(\epsee^{u},\epsee^{d}) \approx (-1,0)$ is within
the $1\sigma$ region.

Let us therefore adopt the hypothesis of NSI with up quarks only and
check whether solar neutrino and CHARM data could break the mass
ordering degeneracy for DUNE. The contour curves in
Fig.~\ref{fig:Vplot} show the combined analysis, where we include
global oscillation data (including the solar neutrino ``SNO-POLY''
analysis) from \cite{Gonzalez-Garcia:2013usa}, assuming that
$\epsee^u$ and $\epset^u$ are approximately uncorrelated.  If we
restrict $\epset = 0$, the degeneracy is broken, since $\epsee^u = -1$
(as required by solar data) implies $\epsee^\oplus \approx -3$,
which can be excluded at high confidence level by DUNE for $\epset =
0$. However, if we allow for non-zero $\epset$, we see that a
large region with the flipped mass spectrum remains below the
$2\sigma$ level around $\epsee^\oplus \approx -3$ and
$|\epset^\oplus| \approx 0.2$. Hence, we conclude that 
including present constraints from oscillation and scattering data,
the degeneracy will severely affect the mass ordering sensitivity of
DUNE. Let us note that if more NSI parameters are allowed to vary, the
fit with the flipped spectrum may even improve further.

%%%%%%%%%%%%%%%%%%%%%%%%%%
\section{Conclusions}
\label{sec:conclusions} 
%%%%%%%%%%%%%%%%%%%%%%%%%%

We have demonstrated that the
so-called LMA-dark solution is a manifestation of an exact degeneracy
at the level of the neutrino evolution equation. This
degeneracy makes it impossible to determine the neutrino mass ordering
by neutrino oscillation experiments. It requires NSI comparable in strength to weak interactions. 
The only way to break the degeneracy is via
non-oscillation experiments. We have shown that taking into account
current data on the $\nu_e$ NC cross section excludes NSI needed
for the exact generalized degeneracy (subject to some model dependence); however, the degeneracy remains
to be present at an approximate level, still destroying the mass
ordering sensitivity of planned experiments.  In order to break the
degeneracy at high confidence level, improved data on $\nu_e$ NC
interactions is mandatory. These may be provided, for instance, by coherent neutrino--nucleus 
interaction experiments \cite{Akimov:2015nza, Wong:2008vk,Aguilar-Arevalo:2016qen,Barranco:2005yy}.

\textit{Acknowledgments.} We thank Stefan Vogl and David V.~Forero for useful
discussions. Fermilab is operated by 
Fermi Research Alliance, LLC under Contract No. \protect{De-AC02-07CH11359} 
with the United States Department of Energy. This project
has received funding from the European Union's Horizon 2020 research
and innovation programme under the Marie Sklodowska-Curie grant
agreement No 674896.

\subsection*{Note added}

Note that flavour evolution in oscillation
experiments is only sensitive to differences of the diagonal elements
of the Hamiltonian. Therefore, eq.~\eqref{eq:trf-mat} should be replaced by the more general expression:
\begin{equation}\label{eq:NSI-deg-diff}
  \begin{array}{l}
    (\epsilon_{ee} - \epsilon_{\mu\mu}) \to - (\epsilon_{ee} - \epsilon_{\mu\mu}) - 2  \,,\\
    (\epsilon_{\tau\tau} - \epsilon_{\mu\mu}) \to -(\epsilon_{\tau\tau} - \epsilon_{\mu\mu}) \,, \\
    \epsilon_{\alpha\beta} \to - \epsilon_{\alpha\beta}^* \qquad (\alpha \neq \beta) \,.
  \end{array}  
\end{equation}
Indeed, the degeneracy can also be
realized for zero $\epsilon_{ee}$, shifting the $(-2)$-term to the
$\mu\mu$, $\tau\tau$ entries in the potential.  In the above analysis
including CHARM data we have implicitly assumed
$\epsilon_{\mu\mu}\approx\epsilon_{\tau\tau} \approx 0$.  The
assumption $\epsilon_{\mu\mu}\approx 0$ is motivated by strong limits
from NuTeV~\cite{Zeller:2001hh}, which should apply under the
assumptions of heavy mediator particles as relevant for CHARM. Since
oscillation experiments constrain $\epsilon_{\tau\tau} -
\epsilon_{\mu\mu} \approx 0$~\cite{Gonzalez-Garcia:2013usa}, the
combination of oscillation and NuTeV data implies
$\epsilon_{\tau\tau}\approx 0$ as well.  We have performed a detailed
study of the generalized mass ordering degeneracy using a combination
of oscillation plus scattering data in Ref.~\cite{Coloma:2017egw}.
In general, in order to exclude the degeneracy we need
  data from scattering experiments on both, electron-neutrino as well
  as muon-neutrino NC scattering.

%%%%%%%%%%%%%%%%%%%%%%%%%%%%%%%
\bibliographystyle{apsrev}
%\bibliography{list-of-refs}

%
\end{document}